\documentclass[letterpaper, 10 pt, conference]{ieeeconf} \IEEEoverridecommandlockouts                              
\date{}
\usepackage{enumerate, amssymb, comment, amsmath, mathtools, graphics, epsfig}
\usepackage{algorithm, algpseudocode, pifont, dsfont}
\usepackage{multirow}
\usepackage{subcaption, graphicx,epstopdf, multirow}

\newtheorem{definition}{Definition}
\newtheorem{problem}{Problem}
\title{\LARGE \bf
Vulnerability of Fixed-Time Control of Signalized Intersections to Cyber-Tampering
}

\author{Amin Ghafouri\thanks{Emails: amin.ghafouri@vanderbilt.edu, waseem.abbas@vanderbilt.edu, yevgeniy.vorobeychik@vanderbilt.edu, xenofon.koutsoukos@vanderbilt.edu}, Waseem Abbas, Yevgeniy Vorobeychik, and Xenofon Koutsoukos\\Institute for Software Integrated Systems, Vanderbilt University, USA}
\begin{document}
\maketitle
\thispagestyle{empty}
\pagestyle{empty}

\begin{abstract} 
Recent experimental studies have shown that traffic management systems are vulnerable to cyber-attacks on sensor data. This paper studies the vulnerability of fixed-time control of signalized intersections when sensors measuring traffic flow information are compromised and perturbed by an adversary. The problems are formulated by considering three malicious objectives: 1) \textit{worst-case network accumulation}, which aims to destabilize the overall network as much as possible; 2) \textit{worst-case lane accumulation}, which aims to cause worst-case accumulation on some target lanes; and 3) \textit{risk-averse target accumulation}, which aims to reach a target accumulation by making the minimum perturbation to sensor data. The problems are solved using bilevel programming optimization methods. Finally, a case study of a real network is used to illustrate the results.
\end{abstract}

\section {INTRODUCTION}

Recent experimental studies claim that about 200,000 vulnerable traffic control sensors are installed in important cities around the world such as New York, San Francisco, London, and Melbourne \cite{cerrudo:emerging}. This indicates the presence of cyber-threats to traffic management systems, since such systems directly use the data measured by the vulnerable sensors. In order to diminish these threats and design resilient systems, the vulnerability of traffic control systems to cyber-tampering of these sensors must be analyzed as an initial step.

In the traffic management literature, queueing networks are often used to model the movement of traffic \cite{allsop1972estimating, papageorgiou2003review}. For the traffic control purposes, various signal control policies are defined based on the queue length information such as \textit{max-pressure} \cite{varaiya:arbitrary, wongpiromsarn2012distributed}, which is a feedback control policy, and \textit{fixed-time control} \cite{muralidharan2015analysis}, which operates the signal in fixed periodical cycles independent of the traffic state. Although feedback control policies for signalized intersections have advantages in terms of stabilizing the traffic flows, 90 percent of all traffic signals in the US follow fixed-time control policy \cite{koonce:traffic}.

%It is proven that max-pressure policy is maximally stable for such networks. Nonetheless, despite the advantages of feedback control policies for signalized intersections, in the U.S., 90 percent of traffic signals follow fixed-time controls, which operate the signal in a fixed periodic cycle, independent of the traffic state \cite{pedarsani:robust}.

Fixed-time control considers deterministic vehicle flows subject to conservation constraints, constraints on saturation flows, and simultaneous turn movements. The formulation of fixed-time control policy leads to a characterization of feasible demands and fixed-time control with minimum cycle length to accommodate the feasible demands \cite{varaiya2013max}. In this direction, Muralidharan et al. showed that under fixed-time control there is a unique periodic trajectory, which is globally asymptotically stable, that is, every trajectory converges to this periodic trajectory \cite{muralidharan2015analysis}. From the periodic trajectory one can easily calculate possible performance measures such as delay, travel time, amount of service time wasted, and progression quality. 

Owing to the rising strategic risks of cyber-attacks, exploiting vulnerabilities of transportation systems to cyber-attacks has been an active area of research. For instance, recently Cerrudo has shown that wireless sensors can be spoofed to manipulate traffic light timing \cite{cerrudo:emerging}. Similarly, in \cite{ghena2014green}, Ghena et al. analyze the security of traffic infrastructure in cooperation with a road agency located in Michigan. The study reports three major weaknesses in the traffic infrastructure: lack of encryption for the network, lack of secure authentication, and vulnerability to known exploits. Furthermore, Laszka et al. have recently proposed an approach for evaluating vulnerabilities of the transportation network by identifying traffic signals with the greatest impact on congestion \cite{laszka:vulnerability}. They also present that the problem of finding an optimal attack to maximize the congestion is computationally hard, thereby, proposing a polynomial-time heuristic algorithm for computing approximately optimal attacks. Nevertheless, no vulnerability analysis of fixed-time control policy has been done for transportation networks. 

In this paper, we study the vulnerability of fixed-time control when a malicious adversary compromises some sensors and perturbs the data corresponding to the traffic flow information. The attacker launches this integrity attack either by directly compromising sensors or by gaining control over the communication network. The tampered data can lead to inefficient scheduling of traffic signals, and in some extreme cases, it can lead to disastrous congestions. In this direction, we formulate three attack problems: 1) \textit{Worst-case network accumulation}, which aims to destabilize the overall network as much as possible; 2) \textit{Worst-case lane accumulation}, which aims to cause worst-case accumulation on some target lanes; and 3) \textit{Risk-averse target accumulation}, which aims to reach a target accumulation by making the minimum perturbation. We formulate these problems as bilevel programs, in which one optimization problem is embedded within the other. Bilevel programs are intrinsically hard to solve, and even the simplest instance, the linear-linear case, is known to be strongly NP-hard \cite{hansen1992new}. The existing algorithms for solving bilevel programs include branch-and-bound, extreme point, complementary pivot, descent methods, penalty function, and trust-region \cite{colson:overview}. We solve the problems using existing implementations of branch-and-bound. Further, we present a case study of vulnerability analysis of a real road network segment in the city of Nashville.

%Such problems appear commonly in various real-world applications in the domains of decision science, engineering, and economic planning.

The remainder of this paper is organized as follows. Section II defines the system model. In Section III, we present the attacker model and formulate the problems. In Section IV, we discuss how the problems can be solved. Sevtion V presents the case study of vulnerability analysis of a real road network. Finally, we conclude the paper in Section VI with a discussion and future work.

\section{SYSTEM MODEL}

\subsection{Network Model}
We use the network model presented in \cite{varaiya2013max} with minor modifications in notation. Consider a network of roads modeled as a directed graph with road links being edges $i \in \mathcal{L}_{all}$ and intersections being nodes $n \in \mathcal{N}$. A link can be either an internal link ($i \in \mathcal{L}$) that goes from its start node to its end node, an entry link ($i \in \mathcal{L}_{ent}$) that has no start node, or an exit link $i \in \mathcal{L}_{exit}$ that has no end node.

A movement $(i,j)$ describes an intention to travel from a link $i$ to a link $j$. Let the \textit{flow} corresponding to movement $(i,j)$ be denoted by $f(i,j)$. This means the rate of vehicles intending to leave link $i$ and enter link $j$ per sample period is $f(i,j)$.
Flow conservation imposes the following constraint on all $i \in \mathcal{L}$,
\begin{equation}\label{fconservation}
\sum_{h\in In(i)} f(h,i) = \sum_{j \in Out(i)} f(i,j)
\end{equation}
where $In(i)$ and $Out(i)$ are the sets of upstream and downstream links connected to $i$. This represents the same concept as the formulation presented in \cite{varaiya2013max}, with routing proportions being implicit in the formulation of each flow $f(i,j)$. %we exclude routing proportions in our model since $f(i,j)$

%Also, let $f_i$ denote flow of vehicles in a link $i \in \mathcal{L}_{all}$. 

Intersections are modeled as nodes and traffic signals are placed at every node to limit the set of permitted movements. Defining a \textit{phase} as a pair of links with $j \in Out(i)$ and $i \in \mathcal{L} \cup \mathcal{L}_{ent}$, \textit{saturation flow} of phase $(i,j)$ is denoted by $c(i,j)$. This means that if phase $(i,j)$ is activated, up to $c(i,j)$ vehicles can move from $i$ to $j$ per sample period.

\begin{figure}
	\begin{center}
		\includegraphics[width=8cm]{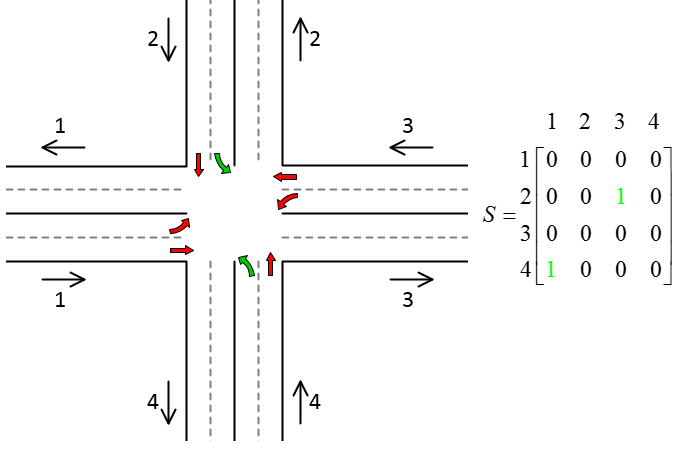}
		\caption{The eight phases of a standard intersection and the control matrix $S$ corresponding to the stage (NW, SE).} 
		\label{fig:S}
	\end{center}
\end{figure}

At an intersection $n$, certain subsets of phases may be simultaneously activated, which is defined as a \textit{stage}. Let $I(n)$ and $O(n)$ denote the set of links entering and leaving intersection $n$. As shown in Fig. \ref{fig:S}, each stage is represented by an intersection control matrix $S_n = \{S_n (i,j), i \in I(n), j \in O(n)\}$ with entries $S_n(i,j) = 1$, if the phase $(i,j)$ is activated, or $0$ otherwise. A collection of intersection control matrices ${S_n}$, one for each intersection, can be combined into the single network control matrix $S$, with $S(i,j) = 1$ if for some intersection $n$, $i \in I(n)$, $j \in O(n)$, and $S_n(i,j) = 1$; otherwise $S(i,j) = 0$. The matrix $S$ can be put in block-diagonal form with the intersection matrices $S_n$ along the diagonal and all other entries zero. The set of all network control matrices $S$ is denoted by $\mathbb{S}$, which is a finite set of $0$, $1$ matrices.
%The set of all such matrices at intersection $n$ is denoted by $\mathbb{S}_n$.

\subsection{Fixed-time Control}
Fixed-time control is a collection of network control matrices $S^1, . . . , S^k$, and corresponding durations $\lambda_{S^1}, . . . , \lambda_{S^k}$, expressed in fractions of a cycle length $T$ \cite{varaiya2013max}. Let $L$ be a fixed lost time per cycle. The minimum cycle length $T$ is defined as $T=\frac{L}{1-(\sum\lambda_{S})} \tau$, where $\tau$ is the sample rate in seconds. Suppose $F = \{ f(i,j) \}$ is a fixed flow matrix. The following linear program (LP) solves the fixed-time control problem
\begin{align}
	\label{eq:fixedtime}
	\begin{split}
		 \min & \sum_{S \in \mathbb{S}} \lambda_{S}
		\\
		 \textrm{s.t. }&  \sum_{S \in \mathbb{S}} \lambda_S c(i,j) S(i,j) \geq f(i,j), \textrm{all } (i,j)
		\\
		& \lambda_S \geq 0, \textrm{ all } \forall S\in\mathbb{S}
	\end{split}
\end{align}
Denote by $\lambda^*$ the minimum value of \eqref{eq:fixedtime}. Flow matrix $F$ is feasible if and only if $\lambda^* < 1$ \cite{varaiya2013max}. The fixed-time LP \eqref{eq:fixedtime} is easily solvable since it decomposes into small linear programs, one per intersection.

\subsection{Example}
\begin{figure}
	\begin{center}
		\includegraphics[width=4.5cm]{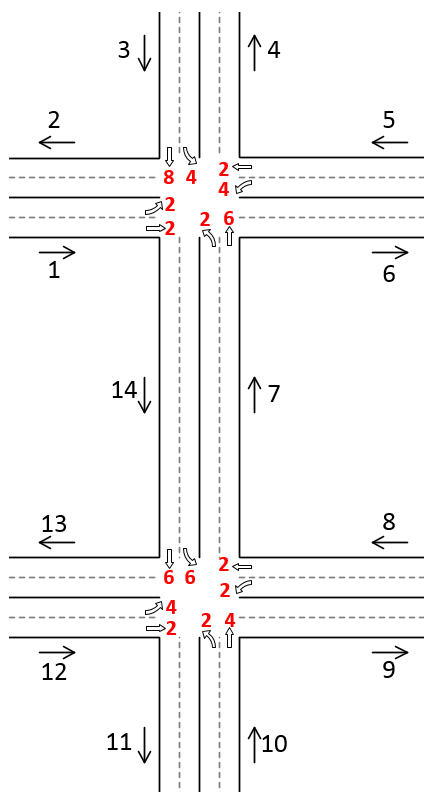}
		\caption{An example of a network with 2 intersections} 
		\label{fig:road}
	\end{center}
\end{figure}

Figure \ref{fig:road} presents a network of 2 intersections with 16 phases. Suppose vehicles flow through the network as the flow data shown in Table \ref{tab0}. Consider four stages (NS,SN), (WE,EW), (NE,SW), and (WN,ES) for each intersection. More specifically, define the stages $\varphi_1=\{(3,14),(7,4)\}$, $\varphi_2=\{(1,6),(5,2)\}$, $\varphi_3=\{(3,6),(7,2)\}$, and $\varphi_4=\{(1,4),(5,14)\}$ for the first intersection, and $\varphi_5=\{(14,11),(10,7)\}$, $\varphi_6=\{(12,9),(8,13)\}$, $\varphi_7=\{(14,9),(10,13)\}$, and $\varphi_8=\{(12,7),(8,11)\}$ for the second intersection.

\begin{table}
	\caption{Flow data used in the example}
	\label{tab0}
	\begin{center}
		\begin{tabular}{ | c | c | c || c | c | c |}
			\hline
			From & To & Flow & From & To & Flow \\ \hline
			\multirow{2}{*}{$1$} & $6$ & $2$ & \multirow{2}{*}{$8$} & $13$ & $2$ \\ \cline{2-3} \cline{5-6}
			& $4$ & $2$ &   & $11$ & $2$ \\ \hline
			\multirow{2}{*}{$3$} & $14$ & $8$ & \multirow{2}{*}{$10$} & $7$ & $4$ \\ \cline{2-3} \cline{5-6}
			    & $6$ & $4$ &    & $13$ & $2$ \\ \hline
			\multirow{2}{*}{$5$}& $2$ & $2$ & \multirow{2}{*}{$12$} & $9$ & $2$ \\  \cline{2-3} \cline{5-6}
			    & $14$ & $4$ &     & $7$ & $4$ \\ \hline
			\multirow{2}{*}{$7$}& $4$ & $6$ & \multirow{2}{*}{$14$} & $11$ & $6$ \\  \cline{2-3} \cline{5-6}
			    & $2$ & $2$ &    & $9$ & $6$ \\ \hline
		\end{tabular}
	\end{center}
\end{table}

Suppose the capacities for all the phases of the first and second intersection are respectively, $32$ and $24$, and let $\lambda = (\lambda_{\varphi_1},\lambda_{\varphi_2},\lambda_{\varphi_3},\lambda_{\varphi_4}, \lambda_{\varphi_5},\lambda_{\varphi_6},\lambda_{\varphi_7},\lambda_{\varphi_8})$. Solving the LP \eqref{eq:fixedtime} for  each intersection, the fixed-time durations are obtained as shown in Table \ref{tab1}. Consequently, for the first and second intersections we obtain $\lambda = \sum_{1}^{4} \lambda_{\varphi_i} = 0.5625$ and $\lambda' = \sum_5^8 \lambda_{\varphi_i} = 0.75$, respectively. Assuming $L=1$, if the same cycle length is required for the entire network, it is computed as $T = \frac{1}{1-\max(\lambda,\lambda')}\tau=4\tau$, where $\tau$ is the sample rate in seconds.

\begin{table}
	\caption{Fixed-time durations}
	\label{tab1}
	\begin{center}
		\begin{tabular}{ |c |c | c | c | c| c| c| c| c| c|}
			\hline
			Stage &$\varphi_1$ & $\varphi_2$ & $\varphi_3$ & $\varphi_4$ & $\varphi_5$ & $\varphi_6$ & $\varphi_7$ &  $\varphi_8$ \\ \hline
			Duration & .25 & .062 & .125 & .125 & .25 & .083 & .25 & .166   \\ \hline
		\end{tabular}
	\end{center}
\end{table}

\section{ATTACKER MODEL}
In this section, we provide a formulation for attacker models that could result in congestion on road networks implementing fixed-time control policy. 
We assume that the attacker knows the network model, fixed-time algorithm, implementation, and can thus compute the optimal schedule. %Further, we assume the adversary has a malicious goal of causing congestion in the network. 

\paragraph{Action Space}
The attacker compromises some of the sensors measuring flows and perturbs their data. Formally, it selects a subset $\tilde{Q}$ of sensors and perturbs their flow values to $\tilde{F}$. Note that we assume the attacker cannot directly change the schedule. But, this can be done indirectly through perturbing the sensor data. This assumption is realistic since it complies with the real-life cyber-attacks launched in the previous experimental studies \cite{cerrudo:emerging}.

\begin{comment}
\begin{figure}
	\begin{center}
		\includegraphics[width=8.4cm]{fig-traffic}
		\caption{Sensor attack on flow sensors} 
		\label{fig:attack}
	\end{center}
\end{figure}
\end{comment}

\paragraph{Objective}

To define the attacker's objective, we first define the notion of a movement being unstable. %Service rate is defined as the total green time of a movement times its saturation flow.
\begin{definition}
	\textit{Unstable Movement}: A movement $(i,j)$ is unstable if its service rate, i.e., $\sum \lambda_S c(i,j) S(i,j))$, is lower than its flow rate $f(i,j)$.
\end{definition}

We assume the adversary's objective is to make some movements unstable, which in turn leads to the network becoming unstable. More specifically, we consider the following different strategies for the adversary:
\begin{enumerate}
	\item \textit{Worst-case network accumulation} which aims to destabilize the overall network as much as possible;
	\item \textit{Worst-case lane accumulation} which aims to cause worst-case accumulation on some target lanes;
	\item \textit{Risk-averse target accumulation} which aims to reach a target accumulation by making the minimum perturbation.
\end{enumerate}

\paragraph{Constraints}
We assume the attacker is resource-bounded, which means that there exists a budget $B$ such that the number of compromised sensors $|\tilde{Q}|$ is less than or equal to $B$, i.e., $|\tilde{Q}| \leq B$. Further, we assume the sensor data and the resulting schedules can only be changed to valid values since otherwise the attack can easily be detected. This means that first, the flow conservation \eqref{fconservation} must be satisfied, and second, the schedule obtained using perturbed data must be feasible, i.e., $\lambda^* < 1$. We formulate the attacker problems assuming traffic signals are timed according to the optimal fixed-time schedule. 

%Note that in practice, to increase fairness, there may exist a minimum green time $g_S \in \mathbb{R}$ for each stage such that $ \lambda_S = g_{S\{\lambda_S \leq g_S\}}$. The attacker may take this observation into account as well when computing the accumulation rate.
%Note that a worst-case attack may lead to perturbed flow value of $0$ for a particular movement if it also satisfies flow conservation. This means if all perturbed flow values of the phases corresponding to a particular stage are $0$, there will be no service time for that stage. However, in practice, this is avoided by considering a minimum green time constraint (i.e., $g_S \in \mathbb{N}$,  $\lambda_S \geq g_S$)
%Further, we assume that due to hardware-based failsafes and fault detectors,

\subsection{Worst-Case Network Accumulation Attack}
The attacker's goal here is to destabilize the network as much as possible and to cause the worst possible traffic congestion. An attack $\mathcal{A}$ has two components of selecting a subset of sensors $\tilde{Q}$ and choosing flow perturbation values $\tilde{F}$. The problem is formally defined below.

\begin{problem}\label{wc}
	\textit{Worst-case Network Accumulation Attack}: Given a network of signalized intersections and a budget $B$, find a worst-case attack $\mathcal{A}=(\tilde{Q},\tilde{F})$ such that it minimizes the service rate of the entire network.
\end{problem}

This problem can be formulated as the bilevel program below.
\begin{align}\label{eq:worst-case}
	\begin{split}
		\max_{\tilde{Q},\tilde{F}}\quad &\sum_{ij} \max(0, (f_{ij} - \sum_S \tilde{\lambda}_S c_{ij} S_{ij}))  \\
		\textup{s.t.}\quad&\tilde{\lambda}_S \in \textrm{FT}(\tilde{F}) \\
		&\sum \tilde{\lambda}_S  <  1 \\
		&\sum_{h} \tilde{f}(h,i) = \sum_j \tilde{f}(i,j) \\
		& |\tilde{Q}| \leq B \\
		&\tilde{f}(i,j) \geq 0, \textrm{ all } (i,j)
	\end{split}
\end{align}

In the formulation above, the term $f_{ij} - \sum_S \tilde{\lambda}_S c_{ij} S_{ij}$, describes the difference between the flow and the service rate. The malicious attacker is only concerned with the positive values for this difference, since negative difference means extra service time, which indeed results in no accumulation. Therefore, the max function is used in the objective function to avoid the negative differences. The first constraint represents the inner-level problem, where $\textrm{FT}(\tilde{F})$ corresponds to the fixed-time LP \eqref{eq:fixedtime} with flow matrix $\tilde{F}$ as its input. The other constraints represent the feasibility of schedule, flow conservation, and attacker's budget respectively.

\subsection{Worst-Case Lane Accumulation Attack}
In a targeted attack, the attacker's goal is to maximize the accumulation rate of a particular lane, or similarly to minimize its corresponding service rates, as much as possible.

\begin{problem}\textit{Worst-Case Lane Accumulation Attack}: Given a network of signalized intersections, budget $B$, and a target lane $l^a$, find an attack $\mathcal{A}=(\tilde{Q},\tilde{F})$  that minimizes the service rate of movements corresponding to the lane $l^a$.
\end{problem}

This problem is formulated as 
\begin{align}
\begin{split}
\min_{\tilde{Q},\tilde{F}}\quad& \sum_j \sum_S \tilde{\lambda}_S c(l^a,j) S(l^a,j) \\
\textup{s.t.}\quad&\tilde{\lambda}_S \in \textrm{FT}(\tilde{F}) \\
&\sum \tilde{\lambda}_S  <  1 \\
&\sum_{h} \tilde{f}(h,i) = \sum_j \tilde{f}(i,j) \\
& |\tilde{Q}| \leq B \\
%&f(l^a,m) \geq \sum_S \tilde{\lambda}_S c(l^a,m) S(l^a,m)\\
&\tilde{f}(i,j) \geq 0, \textrm{ all } (i,j)
\end{split}
\end{align}

The objective function is defined as the sum of the service rates of all movements starting from $l^a$. By minimizing this function, the target lane $l^a$ will have a minimum service time. Note that similar to the previous case, the attacker is restricted by the feasibility of schedule, flow conservation, and budget constraint.

\subsection{Risk-Averse Target Accumulation Attack}
A risk-averse attacker has the strategy of reaching a target accumulation rate while minimizing the perturbations. That is, the difference between the perturbed and actual flow values (i.e., $\lVert \tilde{F} - F \rVert$) must be minimal.

\begin{problem}\textit{Risk-Averse Target Accumulation Attack}: Given a network of signalized intersections, find an attack $\mathcal{A}=(\tilde{Q},\tilde{F})$ that leads to an unstable service rate of $\{\alpha_{ij} \}$ for some set of target movements $Q^a = \{(i,j)\}$, via causing a minimum perturbation $\lVert \tilde{F} - F \rVert$. 
This problem is formulated as the optimization problem below.
\begin{align}
\begin{split}
\min_{\tilde{Q},\tilde{F}}\quad& \lVert \tilde{F} -F \rVert_\infty \\
\textup{s.t.}\quad&\tilde{\lambda}_S \in \textrm{FT}(\tilde{F}) \\
&\sum_S \tilde{\lambda}_S c(i,j) S(i,j) \leq \alpha_{ij}, \forall (i,j) \in Q^a \\
&\sum \tilde{\lambda}_S  <  1 \\
&\sum_{h} \tilde{f}(h,i) = \sum_j \tilde{f}(i,j) \\
&|\tilde{Q}|  \leq B \\
&\tilde{f}(i,j) \geq 0, \textrm{ all } (i,j)
\end{split}
\end{align}
\end{problem}\hfill\\
Note that any other desired norm function can also be used in the objective function.

\section{VULNERABILITY ANALYSIS}
In this section, we present solution and evaluation methods followed by an example.
%Note that the development of efficient algorithms and heuristics for solving the above problem can be part of the challenge in analyzing the vulnerability of a system.

\subsection{Solution}

The three problems described above are all strongly NP-hard, but can be solved with the using of integer programming and decomposition algorithms \cite{colson:overview} \cite{vicente1994bilevel}. Although the computational results and finer details of these algorithms have been suppressed here due to space limitations, we discuss some preprocessing steps carried out in order to be able to use known algorithms for solving bilevel programs.

\subsubsection{Preprocessing} 
In order to handle the max function in the objective function of the first problem, one can convert the problem to a bilevel mixed-integer quadratic program (BMIQP) as follows. In the objective function, for each term of the form $\max(0, (f_{ij} - \sum_S \tilde{\lambda}_S c_{ij} S_{ij}))$, we introduce an auxiliary binary variable $y_{ij} \in \{0,1\}$, and add the constraint $f_{ij} - \sum_S \tilde{\lambda}_S c_{ij} S_{ij} \leq M y_{ij}$, where $M$ is a sufficiently large constant. Then, we replace the previous objective function with:
\begin{align}
\begin{split}
\max_{\tilde{F}}\quad &\sum_{ij} (y_{ij} f_{ij} - y_{ij} \sum_S \tilde{\lambda}_S c_{ij} S_{ij}) \\
\textup{s.t.}\quad
&f_{ij} - \sum_S \tilde{\lambda}_S c_{ij} S_{ij} \leq M y_{ij}\\
\end{split}
\end{align}
In order to solve the risk-averse attacker problem, we rewrite the objective function $\min \; \lVert \tilde{F} - F \rVert_\infty$ as
\begin{align}
\begin{split}
\min \quad& y  \\
\textup{s.t.}\quad& -y \textbf{1} \leq \tilde{F} - F \leq y\textbf{1}
\end{split}
\end{align}
where \textbf{1} is a vector of ones. 

 %We use the following algorithm to solve these problems:

\subsubsection{Solver} The problems are solved using methods for solving bilevel mixed integer programs. Existing algorithms in the literature include branch-and-bound, cutting planes, etc. \cite{colson:overview}. We use the optimization solver Gurobi to solve the attacker problems \cite{gurobi}. We use the MATLAB toolbox YALMIP to invoke Gurobi's bilevel solver \cite{lofberg2004yalmip}. Also, note that because of the worst-case nature of the first and second problems, the optimal value of corrupted flow has to be as small as possible, and thus, the solver can skip considering different values of $\tilde{F}$ and only try extreme values.

\begin{figure*}
	\centering
	\begin{subfigure}{0.315\textwidth}
		\includegraphics[width=\linewidth]{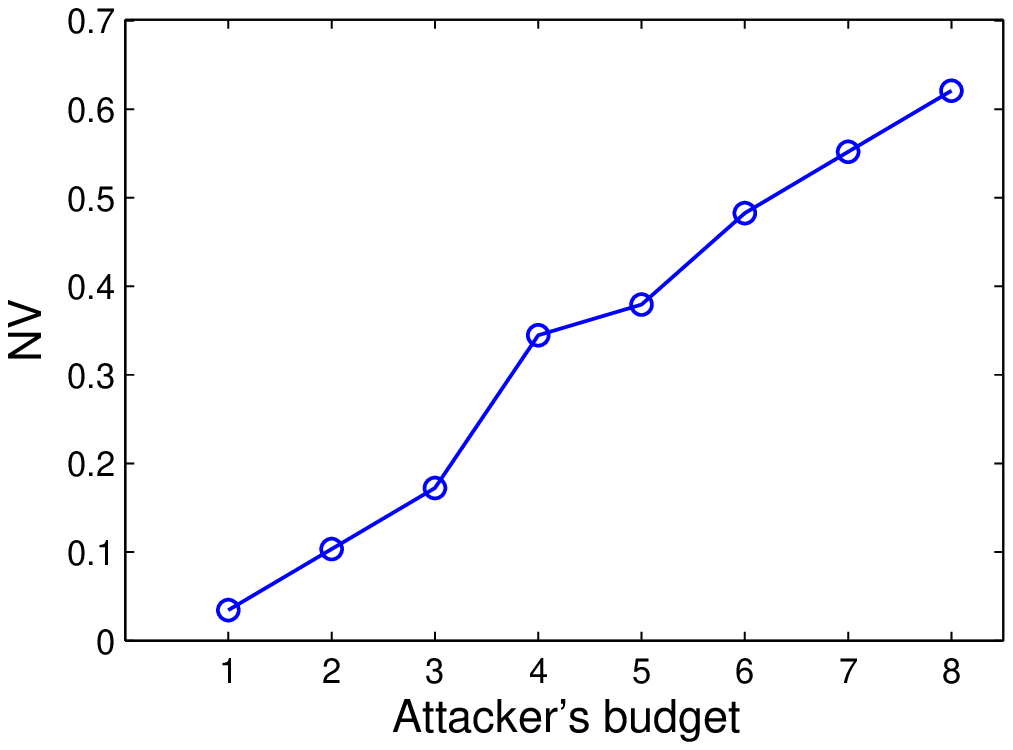}
		\caption{} \label{fig:basicworst}
	\end{subfigure}
	~
	\begin{subfigure}{0.315\textwidth}
		\includegraphics[width=\linewidth]{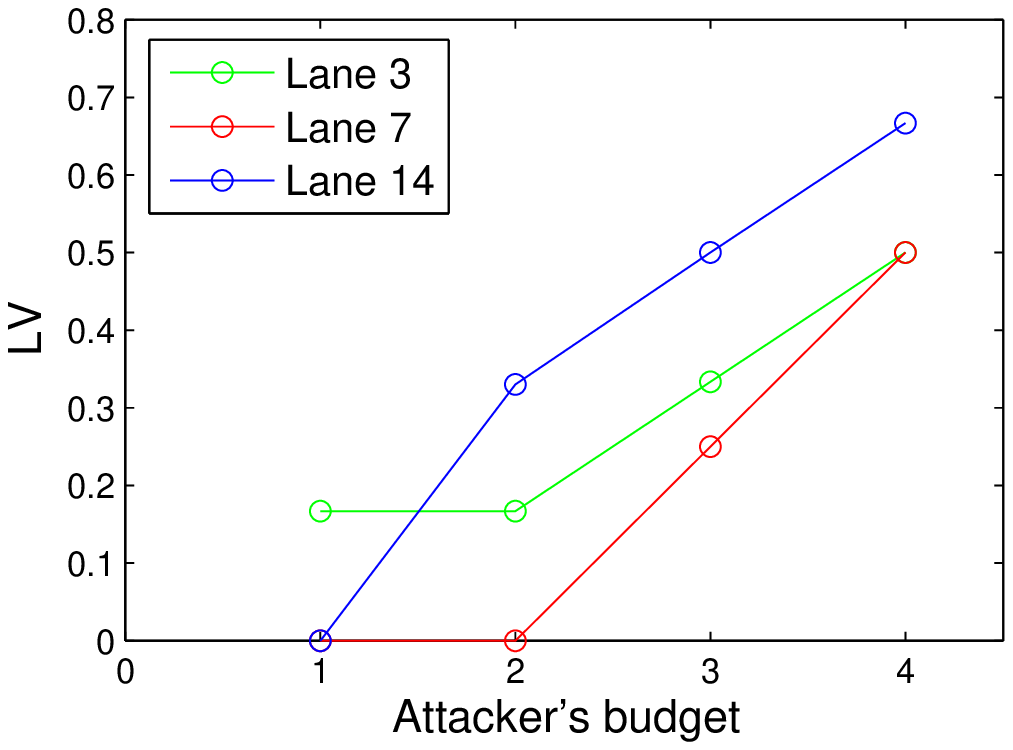}
		\caption{} \label{fig:basictarget}
	\end{subfigure}
	~
	\begin{subfigure}{0.315\textwidth}
		\includegraphics[width=\linewidth]{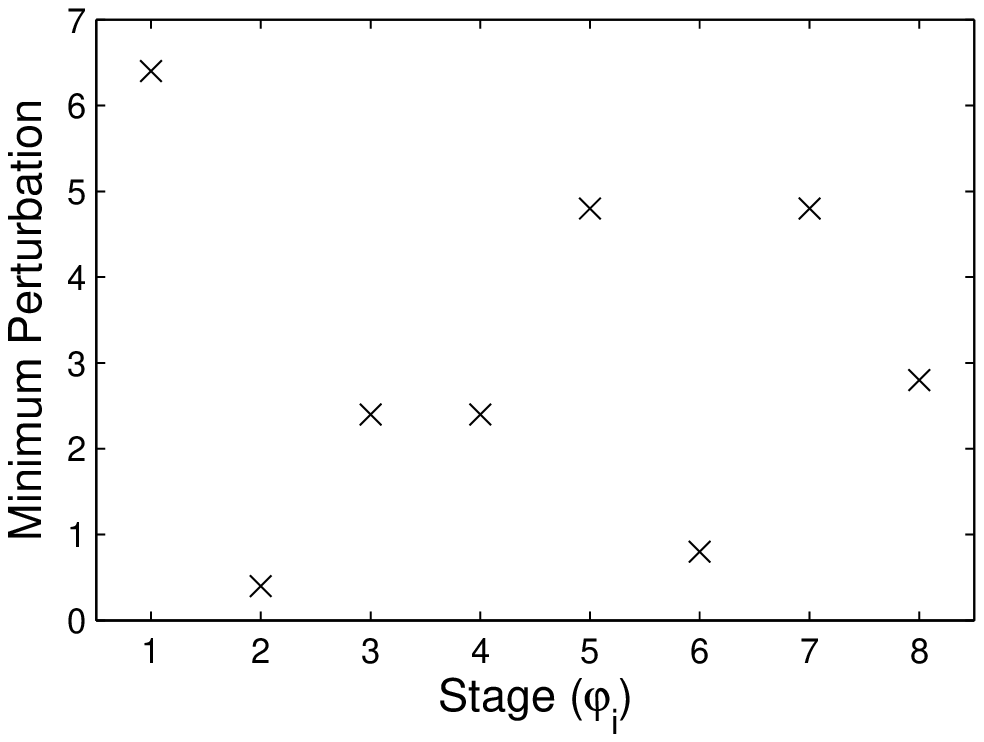}
		\caption{} \label{fig:basicrisk}
	\end{subfigure}
	%\hspace*{\fill} % separation between the subfigures
	\caption{(a) Network vulnerability as a function of attacker's budget in the case of worst case network attack. (b) Lane vulnerability as a function of attacker's budget in the case of worst case lane attacks. (c) Risk-averse target attack. The attacker's goal is to reduce the service time fractions to $0.05$ by making the minimum perturbation.}
	\label{fig:sumo}
\end{figure*}
%\subsection{Critical Sensors and Intersections}

\subsection{Evaluation}

\subsubsection{Metrics}
In order to quantify the vulnerability to worst-case network accumulation attack, we define the network vulnerability as follows:
\begin{definition}
	\textit{Network Vulnerability}: The vulnerability of a network to cyber-tampering is
\begin{equation} 
NV = \frac{\textrm{Accumulation Rate}}{\textrm{Total Flow}}  
\end{equation}
\end{definition}
\hfill\\ 
In the definition above, accumulation rate is the total difference between traffic flow and service rate, i.e., $\sum_{ij} \max(0, (f_{ij} - \sum_S \tilde{\lambda}_S c_{ij} S_{ij}))$, and total flow is the sum of all flow values, i.e., $\sum_{ij} f(i,j)$. The value of network vulnerability represents the relative traffic congestion caused by an attack. We also define lane vulnerability as follows:
\begin{definition}
	\textit{Lane Vulnerability}: The vulnerability of a lane to cyber-tampering is
\begin{equation} 
 LV = \frac{\textrm{Lane Accumulation Rate}}{\textrm{Lane Total Flow}}
\end{equation}
where similarly, lane accumulation is the difference between flow and service rate of the lane, and lane flow is the sum of its corresponding flow values.
\end{definition}

\subsubsection{Critical Sensor}
Besides quantifying the vulnerability of network and lanes, we define critical sensors, which have the highest effect on congestion, as follows:
\begin{definition}
	\textit{Critical Sensor}: A sensor is critical with respect to an attacker's strategy, if it is included in the worst-case attack.
\end{definition}

%Note that critical sensors make natural targets for attacks. 
Identifying the critical sensors allows us to locate the most vulnerable elements of a network, which should be strengthened first to increase the network's resilience. For instance, if there is a security budget that permits us to replace only a subset of the sensors with more secure ones, then we should start with replacing the critical sensors.

%Critical intersection is defined as below.
%\begin{definition}
%	\textit{Critical Intersection}: An intersection is critical if it has the highest number of critical sensors.
%\end{definition}

%Our approach enables us to identify critical traffic signals, which have the greatest impact on traffic congestion and which, therefore, make natural targets for attacks. Identifying these critical signals is beneficial, since it allows us to locate the most vulnerable elements of a network, which should be strengthened first to increase the resilience of a network. For example, if we have a limited security budget which permits us to replace only a subset of the traffic signals with more secure ones, then we should start with the critical signals $\tilde{S}$.
%A traffic signal $s$ is critical if $s \in \tilde{S}$ for a worst-case attack $A$. 

\subsection{Example}

We now study the attacker problems for the network of Fig. \ref{fig:road}. We first solve the worst-case accumulation rate problem \eqref{eq:worst-case}. The results are shown in Fig. \ref{fig:basicworst} as a function of the attacker's budget $B$. As the budget increases (i.e., the attacker is able to compromise more sensors), the accumulation rate increases as well. Also, the results indicate that by controlling only 4 sensors, the attacker can decrease the total service time by up to 35\%.
\begin{comment}
\begin{figure}
	 \centering
	\includegraphics[scale=0.65]{fig-graph1}
	\caption{Network vulnerability as a function of attacker's budget in the case of worst case network attack.} 
\end{figure}
\end{comment}

%By attacking only 4 sensors, the attacker can cause an accumulation ratio of 35\%.

The worst-case lane accumulation problem is solved similarly. Fig. \ref{fig:basictarget} shows the results for the lanes 3, 7, and 14 as targets according to different budgets.
\begin{comment}
\begin{figure} \centering
	\includegraphics[scale=0.65]{fig-graph2}
	\caption{Lane vulnerability as a function of attacker's budget in the case of worst case lane attacks.} 
\end{figure}
\end{comment}
Finally, for the case of risk-averse target accumulation attacks, assume the attacker's objective is to find the minimum perturbation that leads to the target service rate of $0.05$, which is indeed unstable for any stage. Fig. \ref{fig:basicrisk} shows the minimum perturbation for each stage. 

\begin{comment}
\begin{figure} \centering
	\includegraphics[scale=0.65]{fig-graph3}
	\caption{Risk-averse target attack. The attacker's goal is to reduce the service time fractions to $0.05$ by making the minimum perturbation.}
\end{figure}
\end{comment}

%Suppose $B=2$ and $l_{13}$ is the target. It can be shown that no feasible attack exists for this scenario. Next, suppose $B=3$, solving the worst-case attack optimization problem we obtain $\tilde{F}(13,8)=2$, $\tilde{F}(13,10)=2$, $\tilde{F}(2,13)=0$. This leads to the new FT schedules $(1/4,0,1/3,0)$ and $(0,1/4,0,1/12)$, and the service rate of only $\sigma_{13} = 1/12 * 24 = 2$.
\section{CASE STUDY}

We analyze the vulnerability of a real road network segment in the city of Nashville, TN. The area spans between 1st Ave, 8th Ave, Demonbreun St, and Charlotte Ave. The network under consideration comprises 15 intersections (12 four-way and 3 three-way), and 104 movements. In order to perform vulnerability analysis, we use real traffic history data provided by Tennessee Department Of Transportation (TDOT) \cite{tdot}. For lanes with no available data, we estimate their demands using data from their adjacent lanes. Also, since our dataset only provides demands for unidirectional movements, we estimate bidirectional demands considering flow conservation constraints. We assume that fixed-time schedule is computed based on hourly demand, with the total demand being approximately 15000 vehicles per hour.
%Fig. \ref{fig:nash} shows a satellite image of this area.

\begin{figure*}[!t]
	\centering
	\begin{subfigure}{0.3\textwidth}
		\includegraphics[width=\linewidth]{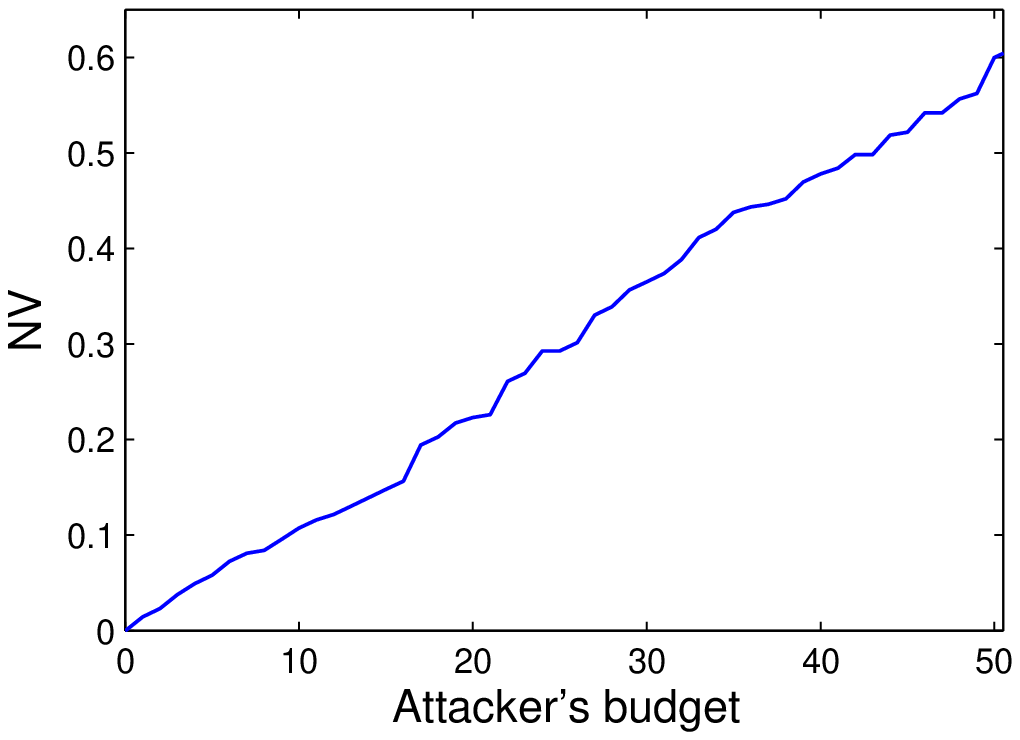}
		\caption{} \label{fig:casegraph1}
	\end{subfigure}
	~
	\begin{subfigure}{0.3\textwidth}
		\includegraphics[width=\linewidth]{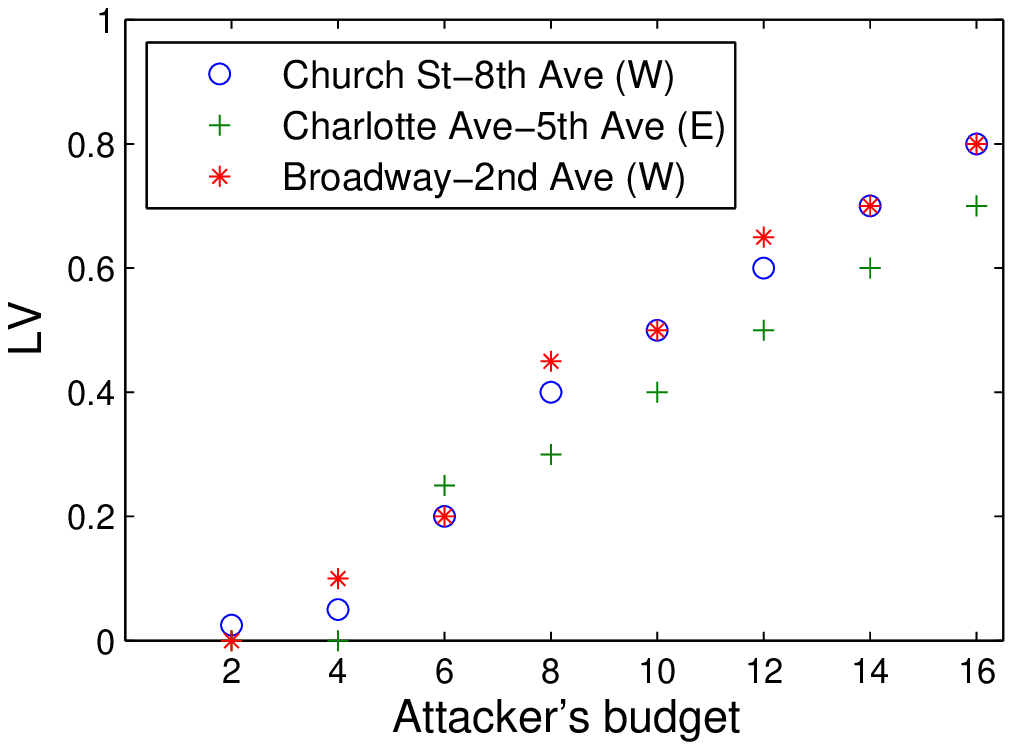}
		\caption{} \label{fig:casegraph2}
	\end{subfigure}
	~
	\begin{subfigure}{0.3\textwidth}
		\includegraphics[width=\linewidth]{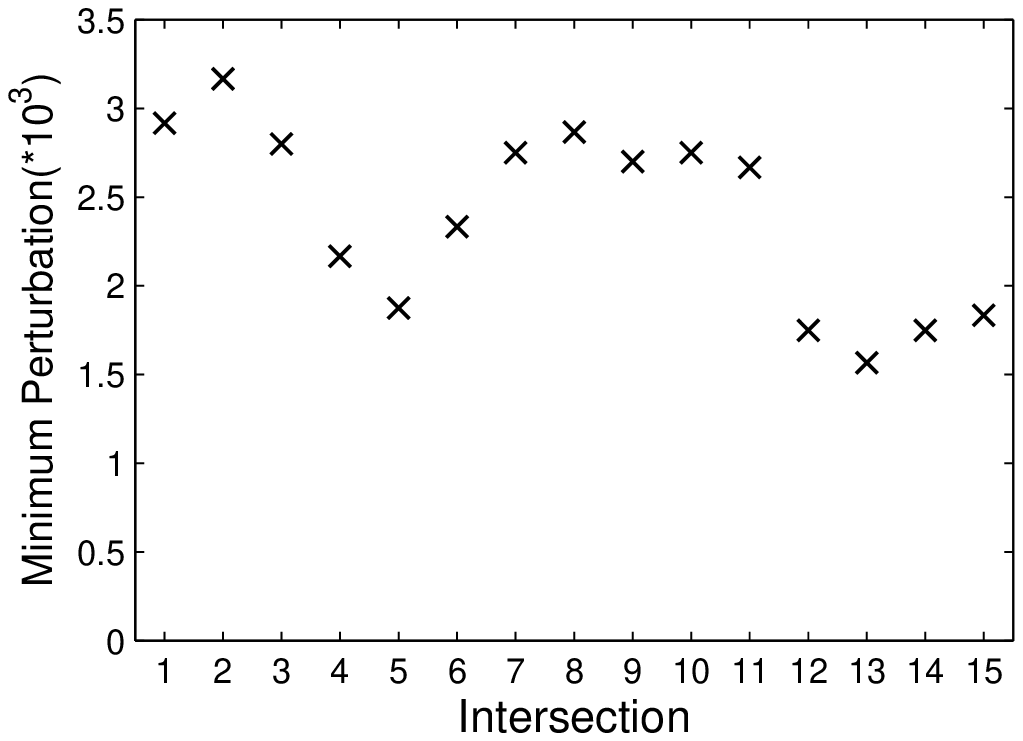}
		\caption{} \label{fig:casegraph3}
	\end{subfigure}
	\caption{(a) Network vulnerability as a function of attacker's budget in the case of worst-case network attack. (b) Lane vulnerability as a function of attacker's budget in the case of worst-case lane attacks. (c) Minimum perturbation needed to reduce the service time of each intersection by at least 50\%, in the case of risk-averse target accumulation attack.}
	\label{fig:case}
\end{figure*}

Figure \ref{fig:casegraph1} presents the results for the worst-case network accumulation problem. The results indicate that by compromising roughly 21 sensors, which is 20\% of the total sensors, the attacker can cause an accumulation of up to 4000 vehicles per hour. Table \ref{table:critical} shows the sensors that appeared most frequently in the worst-case attacks scenarios. %Charlotte Ave-8th Ave (WE) appeared  the most followed by Broadway-8th Ave (NW) and Charlotte Ave-8th Ave (SE).
%Even if we consider a minimum green time fraction of $0.1$ for each stage, the attack can still be disastrous.
\begin{comment}
\begin{figure} \centering
	\includegraphics[scale=.65]{fig-casegraph1}
	\caption{Case study. Worst-case network attack.} \label{fig:nashatt}
\end{figure}
\end{comment}

\begin{table}
	\caption{Sensor measurements with the highest frequency of being attacked.}
	\label{table:critical}
	\begin{center}
		\begin{tabular}{ |c | c | c | c| }
			\hline
			Sensor measurement & Frequency \\ \hline
			
			Charlotte Ave-8th Ave (WE) & $98\%$ \\ \hline
			Broadway-8th Ave (NW) & $97\%$ \\ \hline
			Charlotte Ave-8th Ave (SE) & $95\%$ \\ \hline
			Demonbreun St-8th Ave (NE) & $95\%$ \\ \hline
			Charlotte Ave-5th Ave (WE) & $94\%$ \\ \hline
			Charlotte Ave-3rd Ave (NE) & $94\%$ \\ \hline
			Broadway-8th Ave (WE) & $91\%$ \\ \hline
			Broadway-5th Ave (WE) & $83\%$ \\ \hline
		\end{tabular}
	\end{center}
\end{table}

\begin{figure}
	\begin{center}
		\includegraphics[width=7.2cm]{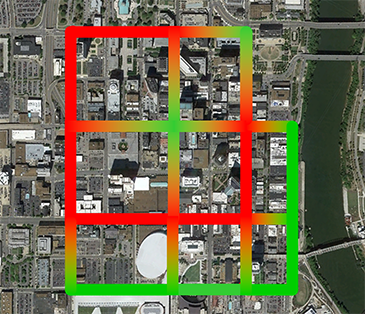}
		\caption{Traffic heatmap of case study after one hour has passed from a worst-case network accumulation attack. Green represents normal traffic and red represents congested traffic.} 
		\label{fig:heat}
	\end{center}
\end{figure}

\begin{comment}
\begin{figure}
	\begin{center}
		\includegraphics[width=8cm]{fig-sathalf}
		\caption{Aerial view of case study. Intersections colored in red are included in at least 50\% of the worst-case network attacks.} 
		\label{fig:sat}
	\end{center}
\end{figure}
\end{comment}
Next, we solve the worst-case lane attack problem for some target lanes. The results are shown in Fig. \ref{fig:casegraph2} for some different budgets. The data shows that on average, it is easier to cause a disastrous congestions on Broadway-2nd Ave than the other two lanes.

\begin{comment}
\begin{figure} \centering
	\includegraphics[scale=.68]{fig-casegraph2}
	\caption{Case study. Worst-case lane attack.}
\end{figure}
\end{comment}
Finally, we solve the risk-averse target attack problem. As the target, we assume the attacker has the goal of reducing the service rate of an intersection by at least 50\%. The results are shown in Fig. \ref{fig:casegraph3} for all 15 intersections. The second and thirteenth intersections (i.e., Charlotte Ave-5th Ave and Demonbreun St-3rd Ave) need the highest and lowest perturbations respectively.

\begin{comment}
\begin{figure} \centering
	\includegraphics[scale=.62]{fig-casegraph3}
	\caption{Case study. Risk-averse target attack for reducing at least 50\% service time of an intersection.}
\end{figure}
\end{comment}

\section{CONCLUSIONS}

We studied the vulnerability of fixed-time control of signalized intersections when sensors measuring traffic flow information are perturbed by an adversary. As the threat model, we considered an attacker that has the objective of congesting the road network. We formulated three attacker problem and solved them using bilevel programming optimization methods. We found that fixed-time control is vulnerable to cyber-attacks and by compromising only a small number of sensors, an attacker can create severe network congestion. Our approach also identified critical sensors, which have the highest impact on congestion. We illustrated our approach by analyzing the vulnerability of a real road network.

%Identifying the critical sensors enables the optimal deployment of defensive countermeasures and resources. 

%Moreover, an attacker could do that while remaining stealthy. etc. etc.

This paper forms the initial step towards more resilient traffic control systems. We aim to extend our results in two directions: first, to design a resilient fixed-time control of signalized intersections so that even if some of the sensors are tampered with, a relatively congestion-free traffic flow is still ensured; and second, to perform the vulnerability analysis of feedback control policies to cyber-tampering.

%We will propose efficient algorithms for finding a resilient configuration, and demonstrate that resilience can be achieved without substantially increasing travel time in the no-attack case. 
%what makes a traffic signal an attractive target by studying the characteristics of critical signals. We will consider basic graph-theoretic metrics (e.g., node degree), characteristics of the traffic flowing through the intersection, and centrality metrics (e.g., betweenness centrality).

\section{ACKNOWLEDGMENTS}
This work is supported in part by the National Science Foundation (CNS-1238959) and the Air Force Research Laboratory (FA 8750-14-2-0180).  Any opinions, findings, and conclusions or recommendations expressed in this material are those of the author(s) and do not necessarily reflect the views of NSF or AFRL.

%\input{control}
%\input{draft}
%\input{mp}

%\nocite{*}
\bibliographystyle{abbrv}
\bibliography{reference}

\end{document}